# Probing spin correlations using angle resolved photoemission in a coupled metallic/Mott insulator system

**Authors:** V. Sunko[1,2]†, F. Mazzola[1]†, S. Kitamura[3]†, S. Khim[2], P. Kushwaha[2], O.J. Clark[1], M. Watson[1], I. Marković[1,2], D. Biswas[1], L. Pourovskii[4,5], T.K. Kim[6], T.-L. Lee[6], P.K. Thakur[6], H. Rosner[2], A. Georges[5], R. Moessner[3], T. Oka[2,3]*, A.P. Mackenzie[1,2]*, P.D.C. King[1]*

**Affiliations:**

[1] SUPA, School of Physics and Astronomy, University of St Andrews, St Andrews KY16 9SS, United Kingdom.

[2] Max Planck Institute for Chemical Physics of Solids, Nöthnitzer Straße 40, 01187 Dresden, Germany.

[3] Max Planck Institute for the Physics of Complex Systems, Nöthnitzer Straße 38, 01187 Dresden, Germany.

[4] CPHT, Ecole Polytechnique, CNRS, Université Paris-Saclay, Route de Saclay, 91128 Palaiseau, France

[5] Institut de Physique, Collège de France, 75005 Paris, France.

[6] Diamond Light Source, Harwell Campus, Didcot, OX11 0DE, United Kingdom.

*Corresponding author. Email: oka@pks.mpg.de (T.O.), Andy.Mackenzie@cpfs.mpg.de (A.P.M.), philip.king@st-andrews.ac.uk (P.D.C.K.)

† These authors contributed equally to this work.

**Abstract:** A nearly free electron metal and a Mott insulating state can be thought of as opposite ends of the spectrum of possibilities for the motion of electrons in a solid. Understanding their interaction lies at the heart of the correlated electron problem. In the magnetic oxide metal $PdCrO_2$, nearly free and Mott-localised electrons exist in alternating layers, forming natural heterostructures. Using angle-resolved photoemission spectroscopy, quantitatively supported by a strong coupling analysis, we show that the coupling between these layers leads to an 'intertwined' excitation that is a convolution of the charge spectrum of the metallic layer and the spin susceptibility of the Mott layer. Our findings establish $PdCrO_2$ as a model system in which to probe Kondo lattice physics, and also open new routes to use the *a priori* non-magnetic probe of photoemission to gain insights into the spin susceptibility of correlated electron materials.

**One Sentence Summary:** An intrinsically non-magnetic spectroscopy is shown to have strong magnetic sensitivity in Kondo-coupled $PdCrO_2$.

**Short Title**: Intertwined spin-charge response of PdCrO2

**Main Text:** $PdCrO_2$ is a member of the broad class of layered triangular lattice materials whose layer stacking sequence (see Fig. 1A) is that of the delafossite structural family $ABO_2$ (*1*). In a simple ionic picture of the delafossites, triangular co-ordinated layers of $A^+$ ions are stacked between oxygen octahedra with $B^{3+}$ ions in the centre, in which the B ions also have triangular co-ordination (*2*, *3*). Most delafossites are insulating or semiconducting. $PdCoO_2$ and $PtCoO_2$, however, are extremely high conductivity metals featuring broad conduction bands whose character is dominantly that of the A site cation Pd or Pt (*4–9*), with the B-site $Co^{3+}$ cation in the band insulating and non-magnetic $3d^6$ configuration.

In contrast, in the Cr-based analogue $PdCrO_2$, the $Cr^{3+}$ cations are formally in the $3d^3$ configuration (Fig. 1B). It was therefore not regarded as surprising when $PdCrO_2$ was observed to be magnetic, obeying a Curie-Weiss law at high temperatures, followed by a transition to 120° antiferromagnetism at a Néel temperature $T_N$ of 37.5 K, carrying a localised spin of S=3/2 (*10–14*). Interestingly, from angle-resolved photoemission (ARPES), Sobota *et al.* found an extremely similar Fermi surface to $PdCoO_2$, indicating that the low energy electronic structure is still dominated by the Pd-derived states (*15*). A reconstruction of the Fermi surface due to the magnetic order was reported by de Haas-van Alphen measurements (*16*, *17*). Careful analysis of magnetic breakdown across the gaps opened at the antiferromagnetic (AF) Brillouin zone boundary showed that these gaps are small, on the order of 40 meV (*16*). Noh *et al.* subsequently reported that the bands are apparently back-folded across the AF zone boundary in ARPES measurements (*18*). Similar spectroscopic signatures are also evident in our own measurements shown in Fig. 1C and D, where weak but clear spectral weight can be observed as replicas of the 'main band' (central hexagonal Pd-derived Fermi surface (*15*)) shifted in momentum by the antiferromagnetic ordering vector.

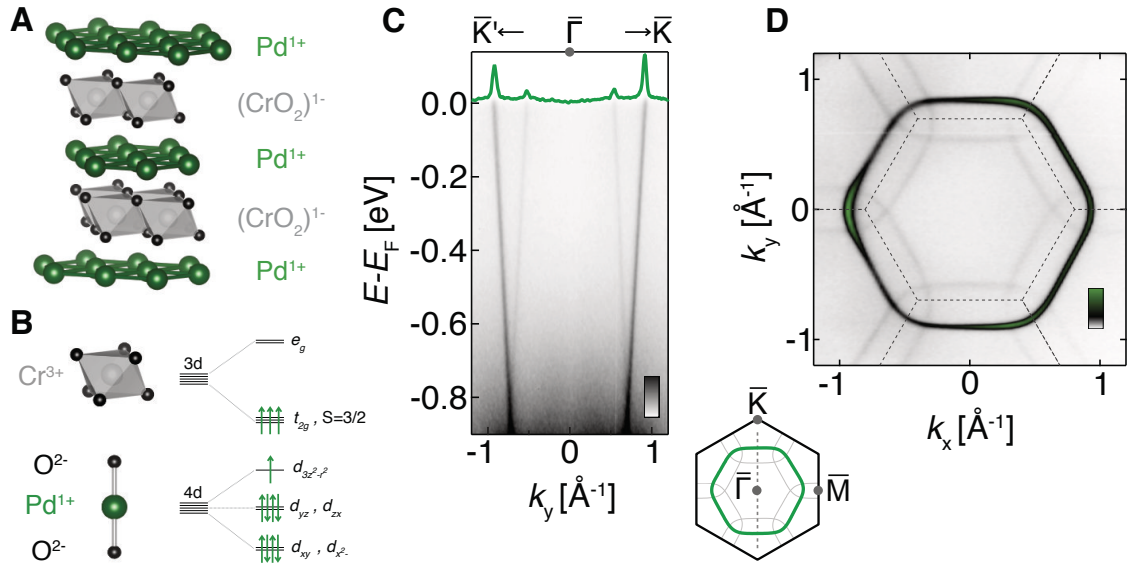

**Fig. 1. Low-energy electronic structure of $PdCrO_2$.** (**A**) The layered crystal structure of $PdCrO_2$. (**B**) Pd layers are metallic, while the $CrO_2$ layers are Mott insulating and antiferromagnetically ordered below $T_N = 37.5$ K. (**C**) Dispersion measured by ARPES ($hv = 110$ eV, $T = 6$ K) along the $\bar{\Gamma}$–$\bar{K}$ direction (dashed line on the schematic of the crystallographic Brillouin zone) showing steep Pd-derived metallic bands, as well as replicas of these bands, apparently back-folded across the magnetic Brillouin zone boundary (dashed lines in D). Strikingly, the observed reconstructed spectral weight is approximately energy independent over nearly 1 eV, remaining clearly visible at the Fermi level, as evident in the momentum distribution curve (green line in C, $E_F \pm 5$ meV), and the measured Fermi surface (**D**, $hv = 120$ eV, $T = 6$ K, integrated over $E_F \pm 25$ meV).

The observation of localised 3/2 spins on the Cr sites strongly suggests that, in addition to being magnetic, the $CrO_2$ layer is Mott insulating (*9, 17, 19*). This hypothesis has recently been supported by combined density functional theory (DFT) and dynamical mean-field theory (DMFT) calculations from our own group (Supplementary Text 2) and independent work by Lechermann (*20*), which concluded that in the paramagnetic state above $T_N$, the conduction in $PdCrO_2$ comes from a single band of dominantly Pd character. The Cr-derived states, predicted by standard DFT to produce additional Fermi surfaces (*20, 21*), instead form a lower Hubbard band giving substantial incoherent spectral weight 1-2 eV below the Fermi level.

In order to experimentally verify this picture, we have used soft X-ray ARPES to investigate the atomically-resolved electronic structure, tuning the probing photon energy into resonance with the Cr $L_{2,3}$ absorption edge (*22*). Comparing on- to off- resonant spectra (Fig. 2A – C) reveals a marked enhancement of spectral weight of a very weakly dispersing and broad feature centred at approximately 2 eV below $E_F$. The integrated intensity of this feature ($I_{LHB}$) tracks the Cr $L_{2,3}$ - edge X-ray absorption spectrum (Fig. 2D), thus establishing its Cr-derived origin. Comparison with the findings from the DFT+DMFT calculations provides strong evidence that this is the spectroscopic signal of a lower Hubbard band. In addition, and consistent with Fig. 1C, we observe a steeply-dispersing metallic band. This band shows negligible change in spectral weight across the Cr $L_{2,3}$ – edge resonances ($I_{MB}$ in Fig. 2D), confirming that it originates from the Pd layers. $PdCrO_2$ can therefore be considered as an atomic layer-by-layer superlattice of a nearly-free electron metal alternating with a Mott insulator.

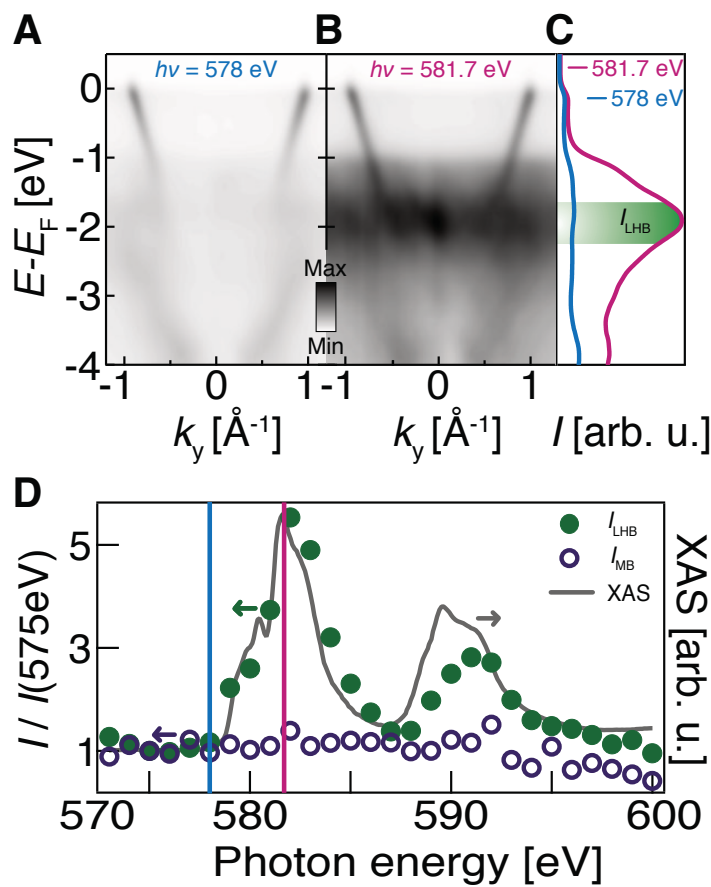

**Fig. 2. Mott insulating $CrO_2$ layers.** Soft X-ray ARPES (T=13K) from $PdCrO_2$ at photon energies of (**A**) 578 eV and (**B**) 581.7 eV, respectively tuned off- and on-resonance with the Cr $L_3$ edge. The on-resonant spectrum reveals considerable broad spectral weight centred at approximately 2 eV below the Fermi level. The measured intensity of this feature, $I_{LHB}$, extracted from energy distribution curves (**C**, integrated over $0\pm0.5$Å$^{-1}$) as a function of probing photon energy is in excellent agreement with the measured X-ray absorption spectrum (XAS) across the Cr $L_{2,3}$ –edge (**D**). In contrast, the intensity of the Pd-derived main band ($I_{MB}$, extracted from fits to momentum distribution curves at the Fermi level) stays approximately constant across the resonance. The data provide strong evidence that the diffuse weight visible in the ARPES measurements is dominantly of Cr character, while comparison with the DFT+DMFT calculations (Supplementary Text 2 and Ref. 20) identifies it as the lower Hubbard band of a Mott insulating state.

Given the antiferromagnetic order of the latter, the observation of replicas of the metallic 'main band' shown in Fig. 1C and D might, at first sight, seem unremarkable. In general, when electrons feel an additional periodic potential, for example due to a density wave or magnetic order, the band structure is reconstructed. Replica bands appear, shifted from the original ones by the wavevectors of the new potential, with hybridisation gaps opening at the new Brillouin zone boundaries. This standard picture, however, cannot explain the experimental observations of $PdCrO_2$. The spectral weight of the replicas observable by ARPES should fall off rapidly away from the new zone boundaries (23), with a form equivalent to the well-known coherence factors of Bogoliubov quasiparticles in a superconductor. Experimentally, however, the replicas are clearly observed all the way from the magnetic zone boundary to the Fermi level (Fig. 1C, D), an energy range more than an order of magnitude larger than the hybridisation gap scale of ~ 40meV (16). Over the same energy range the simple 'band folding' model predicts a 100-fold decrease in spectral weight (dashed line in Fig. 3A), which would render the backfolded bands invisible to ARPES. In contrast, the measured intensity of the reconstructed weight ($I_{RW}$) changes by less than a factor of 2 (symbols in Fig. 3A). Additional measurements performed using different light polarization and photon energies show similarly-weak binding energy dependence of the reconstructed weight (Supplementary Fig. S5); the striking discrepancy with the band folding model cannot, therefore, be explained by photoemission matrix-element variations. A further possibility worthy of investigation is final-state Umklapp scattering, in which an outgoing photoelectron is diffracted from the potential of a superperiodic structure during its travel to the surface. This can in principle yield back-folded bands whose spectral weight has only a weak binding-energy dependence (24-27). In the delafossites, however, this possibility can be ruled out by making a direct comparison with the isostructural, but non-magnetic, $PdCoO_2$. Under identical measurement conditions to those of $PdCrO_2$, we observe no such signal (Supplementary Fig. S3), and so it is clear that the replica features of Figs. 1C and D require a different explanation.

We have discovered that the answer to the above puzzle lies in Mott insulator - free electron coupling. Rather than treating the $CrO_2$ layer as a passive source of a periodic potential, it is necessary to take into account its dynamical degrees of freedom. To illustrate this, we start with a minimal model (shown schematically in Fig. 3B) combining hopping within and between the Pd and $CrO_2$ layers with the Coulomb repulsion in the Mott layer:

$$H = -\sum_{ij\sigma}^{n.n.}(t_p p_{i\sigma}^\dagger p_{j\sigma} + t_c c_{i\sigma}^\dagger c_{j\sigma}) + U\sum_i \left(n_{i\uparrow}^c - \frac{1}{2}\right)\left(n_{i\downarrow}^c - \frac{1}{2}\right) + \sum_{ij\sigma}^{n.n.} g_{ij}(p_{i\sigma}^\dagger c_{j\sigma} + \text{H.c.}), \quad (1)$$

where $t_p$ ($t_c$) denote the hopping integrals between the Pd (Cr) sites, $g$ is the interlayer hopping and $U$ the Coulomb repulsion. We neglect Coulomb repulsion between the Pd electrons, justified by the fast band velocities of the Pd-derived states observed experimentally (see Supplementary Fig. S3C). Here we omit the orbital indices, and assume S=1/2 on the Cr sites for conceptual simplicity.

We present the full multiorbital model with S=3/2, the results of which are shown in Fig. 3D and F, in Supplementary Text 3.

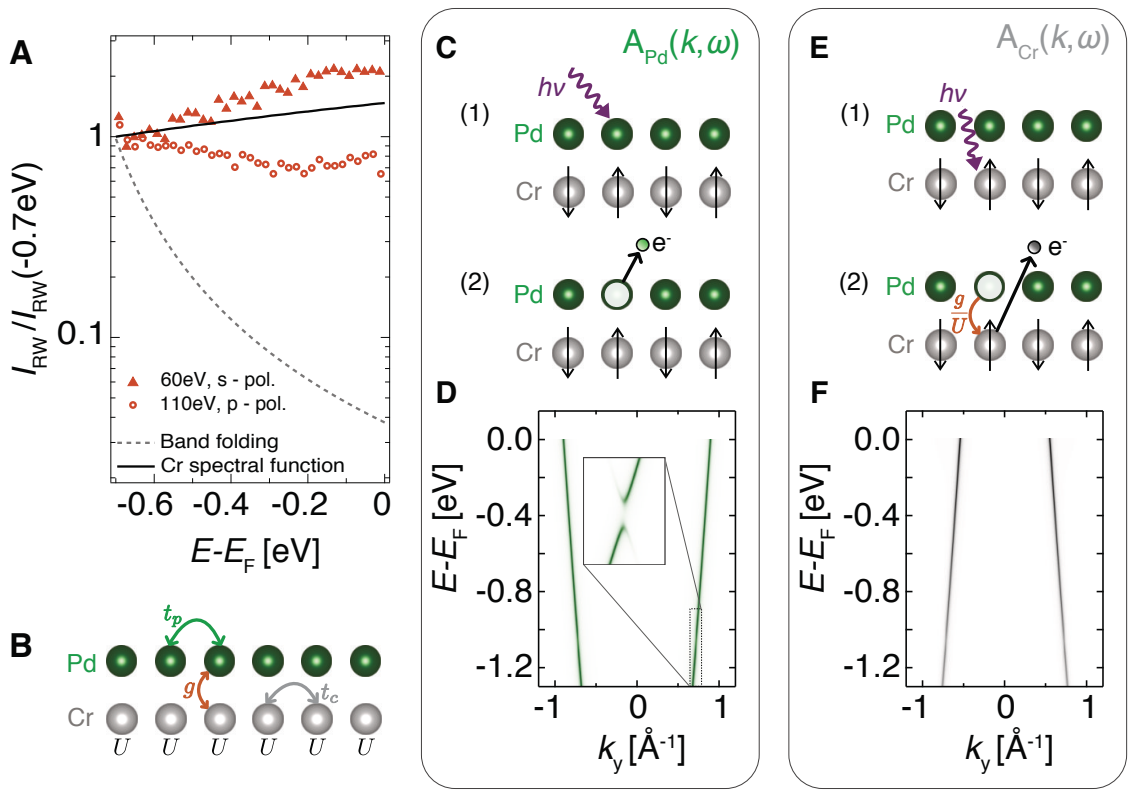

**Fig. 3. Intertwined spin and charge response.** (**A**) Reconstructed weight ($I_{RW}$) as obtained from fits to the dispersion shown in Fig. 1C (circles). Due to photoemission matrix elements small quantitative variations are found when measuring using different photon energies and light polarisations. We also show here data measured using 60eV s-polarised light (triangles) to illustrate the range of observed spectral weight variations; additional measurements are shown in Supplementary Fig. S5. In all cases $I_{RW}$ varies only weakly with binding energy. This is in sharp contrast to the simple 'band folding' model (dashed line, see Supplementary Text 3.2), but in agreement with the Cr spectral function predicted by our theory (solid line, see text). The intensities are shown normalised to the intensity at -0.7 eV binding energy to aid judging the relative binding energy dependent variations in the data and different models; equivalent conclusions are drawn if normalising directly by the main band intensity (Supplemental Fig. S7). (**B**) The starting point of the theory is a Hamiltonian which includes hopping within ($t_p$, $t_c$) and between ($g$) the layers, as well as the on-site Coulomb repulsion on the Cr sites ($U$). (**C**) Schematic illustration of photoemission of Pd electrons. (**D**) The corresponding spectral function is equivalent to that predicted by the 'band folding' model. (**E**) Photoemission of a Cr electron can proceed via a virtual process involving tunnelling of the Cr hole to the Pd layer. (**F**) This results in a spectral function that is a convolution of the Pd spectrum and the spin correlation function of the Mott layer (Equation 3), thus appearing as a copy of the Pd spectral function shifted by the wavevector of the AF order, in agreement with the experiment (Fig. 1C).

The large size of $U$ compared to the other coupling constants enables a standard strong coupling analysis, implemented via a Schrieffer-Wolff transformation (full procedure described in Supplementary Text 3) to derive a low-energy Kondo lattice Hamiltonian:

$$H_{\text{eff}} = -t_p \sum_{ij\sigma}^{n.n.} p_{i\sigma}^\dagger p_{j\sigma} + \frac{4t_c^2}{U} \sum_{\langle ij \rangle}^{n.n.} \boldsymbol{S}_i \cdot \boldsymbol{S}_j + \frac{4}{U} \sum_{ijk\sigma\sigma'}^{n.n.} g_{ij} g_{kj} p_{i\sigma}^\dagger (\boldsymbol{S}_j \cdot \boldsymbol{\sigma}_{\sigma\sigma'}) p_{k\sigma'}, \quad (2)$$

in which the second term captures the effective spin-spin exchange in the Mott layer and the last term describes a Kondo coupling between the localised Cr spin and Pd electrons on the neighbouring sites.

The Hamiltonian (2) is of a standard form, but applied here to an unusual situation in which the Kondo coupling is an interlayer effect. The key insights it provides comes from using it to calculate the spectral functions for the photoemission process. The coupling allows the Pd electrons to feel the periodic potential due to the AF order of the Cr spins, but does not otherwise affect their basic itinerant nature. The resulting Pd one-electron removal spectral function (Fig. 3C, D) thus, unsurprisingly, looks like the simple `band folding' model introduced above: it largely follows the unperturbed Pd dispersion, with small gaps opening at the magnetic zone boundary as seen by quantum oscillations (*16*), and has a weak, strongly energy dependent weight in the reconstructed band.

In contrast, the removal of electrons from Cr orbitals is drastically altered by the coupling to the Pd layer. It would be impossible to remove an electron from an isolated Mott layer at energies smaller than $U$. However, for finite interlayer coupling $g$, a hole created in the Mott layer can rapidly move to the itinerant layer where it can propagate; formally, the Schrieffer–Wolff transformation leads to an effective real space Cr removal operator of the form $(c_{j\sigma})_{\text{eff}} = \frac{2}{U} \sum_k^{n.n. \text{ of } j} g_{kj}(\boldsymbol{S}_j \cdot \boldsymbol{\sigma}_{\sigma\sigma'}) p_{k\sigma'}$. We note two important features of the transformed operator. Firstly, the process is perturbatively small in $g/U$. Secondly, it provides a connection between the itinerant Pd electrons and Mott spins. This results in the spectral function for the removal of electrons from the Mott layer becoming a convolution of the itinerant electron spectrum with the spin correlation function of the Mott layer:

$$A_{\text{Cr}}(\boldsymbol{k}, \omega < 0) = -\int_{-\infty}^0 \frac{d\omega'}{2\pi} \int \frac{d^3\boldsymbol{q}}{(2\pi)^3} \frac{32|g_{\boldsymbol{k}+\boldsymbol{q}}|^2}{U^2} A_{\text{Pd}}(\boldsymbol{k}+\boldsymbol{q}, \omega') \langle \boldsymbol{S}_{\boldsymbol{q}} \cdot \boldsymbol{S}_{-\boldsymbol{q}}(\omega - \omega') \rangle. \quad (3)$$

In this way, the spin response of the Mott layer and the charge response of the itinerant layer become intertwined. In the case of AF ordered PdCrO$_2$, the mean-field spin correlation function is a delta function at zero energy and the AF wavevector. The resulting prediction (Fig. 3E, F) is that Cr spectral weight now exists at energies much lower than $U$, and that it follows the dispersion of the nearly free electron Pd band but translated by the wavevector of the AF order.

This calculation of intertwined spin-charge response yields at least two testable predictions for the observable spectral properties which are qualitatively different from those of a standard band folding model. Firstly, as discussed above, standard band folding predicts a spectral weight of the replica band that dies off extremely quickly with energy away from the magnetic Brillouin zone boundary. In contrast, the intertwined spin-charge model has no inherent energy dependence of the reconstructed weight; momentum dependence of the interlayer coupling constant $g$ can still give small system-specific variations (Supplementary Text 3.2 and Supplementary Fig. S6), but in general the energy dependence of the reconstructed weight will be weak. As shown in Fig. 3A, observation is in close agreement with the latter prediction. The second and even more striking prediction is that the back-folded spectral weight, despite appearing as a sharp band-like feature, is actually a property of the Cr removal spectral function rather than the Pd one. This is at first sight very surprising, because 'Mott electrons' are associated with the broad incoherent spectral weight displayed and described in Fig. 2.

The key diagnostic for the validity of the intertwined spin-charge model is therefore to establish the underlying atomic origin of the reconstructed spectral weight. To do this we again use soft X-ray ARPES to show that the reconstructed weight ($I_{RW}$) is markedly enhanced when the photon energy is tuned to the Cr $L_3$-edge resonance (Fig. 4A-C). Moreover, quantitative analysis of measurements performed at lower photon energies shows (i) that the photon energy dependence of the reconstructed weight closely traces that of the Cr-derived lower Hubbard band (Fig. 4D), and (ii) that its ratio to the Pd-derived main band intensity tracks the Cr $3d$ : Pd $4d$ ionic cross-section ratio (Fig. 4E). These observations all point to a dominant Cr character of the back-folded spectral weight.

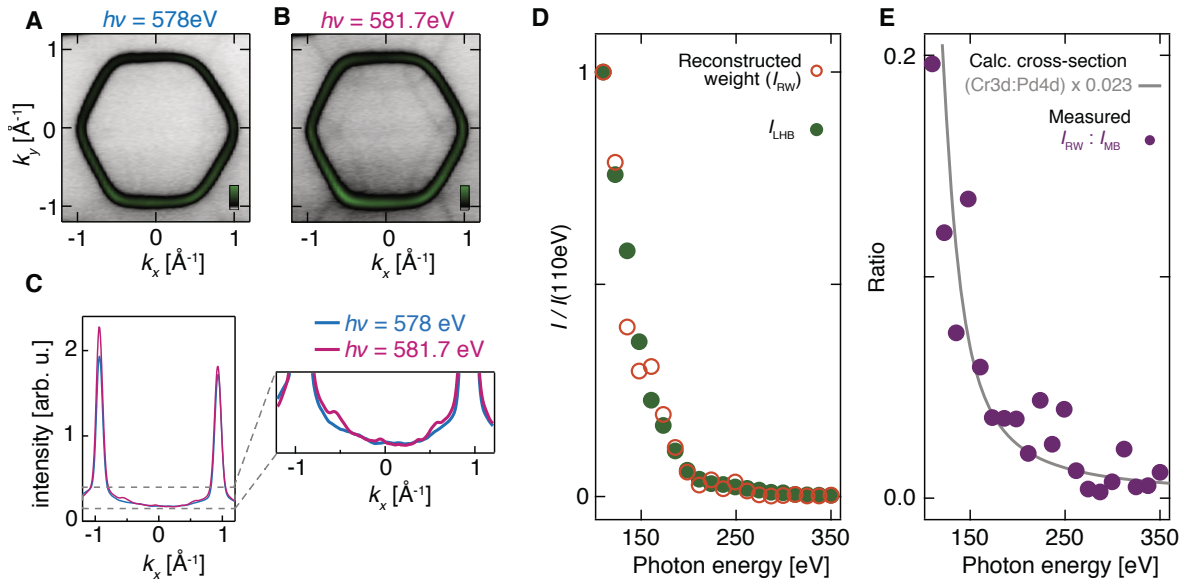

**Fig. 4. Cr origin of the reconstructed weight.** The Fermi surface measured (T = 13 K, integrated over $E_F \pm 200$meV) off-resonance (**A**, $h\nu$ = 578 eV) and on-resonance (**B**, $h\nu$ = 581.7 eV) with the Cr $L_3$-edge. The reconstructed weight is markedly enhanced in the resonant condition, as evident in a comparison of MDCs at the Fermi level recorded on and off resonance (**C**). (**D**) The photon energy dependence of the reconstructed weight ($I_{RW}$) at lower photon energies closely tracks that of the Cr-derived lower Hubbard band ($I_{LHB}$). (**E**) The ratio of $I_{RW}$ to the weight of the 'main band' ($I_{MB}$) is strongly photon energy dependent. It follows the functional form expected for the Cr $3d$ : Pd $4d$ ionic cross-section ratio *(30)*, scaled by a factor of ~0.023, the origin of which is the spectral weight suppression factor of $(g/U)^2$ predicted by the intertwined spin-charge model (Equation 3).

Equation (3) for the Cr spectral function suggests that the weight of the reconstructed feature should be suppressed by approximately $32g^2/U^2$ as compared to the weight of the 'main band' in the Pd spectral function. While ARPES matrix elements prevent us from making a direct quantitative measurement of the intrinsic relative weights, comparison with the ionic cross-section ratio shown in Fig. 4D indicates that their ratio is on the order of 1%. With $U = 4$ eV the inter-layer coupling $g$ is thus estimated to be on the order of 100 meV. This is similar to values derived from a density functional theory analysis of inter-layer hopping (Supplementary Table S1), providing a further consistency check on our analysis.

In principle, a standard one-electron Pd-Cr hybridisation could lead to a Cr character of the replica weight. This would, however, be fundamentally incompatible with the lack of binding energy-dependence of its measured spectral weight. Indeed, obtaining measurable Cr weight up to the Fermi level would require hybridisation gaps on the order of ~1 eV to open where the bands cross (*18*), which is completely inconsistent with the measured electronic structure both from ARPES (Fig. 1C,D) and from quantum oscillations (*16*). On the other hand, while a final-state Umklapp process as introduced above could explain the binding energy independent reconstructed weight, it could not explain its Cr-derived character: if it were due to a final-state Umklapp of the Pd-derived main band of either structural or exchange scattering origin, it would exhibit Pd character, and not the Cr character that we observe experimentally. The *combination* of these two key experimental observations (weak binding-energy dependence and Cr character of the reconstructed weight) thus provides compelling evidence that the spectroscopic information obtained from ARPES measurements on PdCrO$_2$ is determined by a Kondo coupling of nearly free electrons in metallic layers with localized electrons in a Mott insulating state in adjacent layers.

This realization is, we believe, exciting for a number of reasons. It establishes PdCrO$_2$ as a benchmark system for studies of the triangular lattice Hubbard model and Mott insulator-free electron coupling. The transport properties of PdCrO$_2$ *(10-14,17)* are of considerable interest in their own right, and invite further theoretical work. In equations (1) and (2) above, and in more detail in Supplementary section S3, we present model Hamiltonians (including experimentally constrained bare parameters) as a foundation for such calculations. However, we believe that our findings are also of considerable relevance across broader fields of research.

The first regards the interplay of localised and itinerant electrons in solids in general, which lies at the heart of the physics of Kondo systems. PdCrO$_2$ lies in the antiferromagnetic metal region of the famous Doniach phase diagram (*28*) because the localised spin is underscreened. Although this means that the formation of heavy fermions is not expected in the current experimental situation, the insights that we have uncovered open new ways to study systems from across the phase diagram, even in the Kondo limit. This would in principle include quasi-3D materials in which the layer-dependent coupling in PdCrO$_2$ is replaced by, for example, a coupling between localised *f*-electrons and itinerant conduction electrons. The limitation in practice will be one of resolution; if the bandwidths become too small and the $k_z$ dispersion too strong, the spin-related signatures will be harder to extract, but such information should exist in the experimental signal. We also stress that the analysis presented above does not rely on the existence of static magnetic order, and could equally well apply to a disordered system with a peaked susceptibility, of the type discussed in the context of cuprate superconductors (29). In that case the width of the feature observed in photoemission would be related to the antiferromagnetic correlation length and time.

In such a dynamical case, because of the way in which magnetic and charge excitations associated with the different subsystems are intertwined during the photoemission of electrons from the Mott subsystem, we can obtain information on both in a single measurement. Excitingly, there may be a regime of coupling in which separated magnon and charge velocities could become observable by ARPES. Our work therefore has a further, conceptually attractive, component in providing a complementary perspective to the physics of spin-charge separation. There, an electron in a solid fractionalises into independently mobile excitations carrying its magnetic and charge quantum numbers. Another interesting avenue for future theoretical research is to investigate how the existence of a Mott state which is itself fractionalised would be manifested in this type of experiment.

Our findings thus point to the broad potential of combining itinerant and Mott insulating systems in heterostructure geometries, where our analysis indicates that finite coupling between the layers leads to new physics not present in either of the spatially-separated subsystems alone. Consistent with this, a recent DMFT study (20) found that doping of the Mott layer in $PdCrO_2$ results in an effective Pd layer doping. Although the underlying process is not the same as in the photoemission problem studied here, both effects reflect the fact that in a coupled Mott/itinerant system it is the itinerant layer which will support charge excitations.

The insights gained here therefore have relevance to the study of materials well beyond $PdCrO_2$. For example, while the development of spin-resolved detectors has opened new opportunities to study spin-polarised itinerant bands, ARPES is typically not expected to be sensitive to finite $q$ local moment magnetism: our findings show that this need not be true. Our work thus opens an entirely new route to investigating both static and dynamical spin susceptibilities in correlated solids, including systems which are inaccessible to more traditional magnetic probes such as neutron scattering. Combining ARPES studies with targeted materials design of coupled Mott-metal heterostructures could therefore provide unique information on magnetic ordering and fluctuations in previously unexplored regimes in, for example, transition-metal oxide heterostructures, two-dimensional van der Waals magnetic insulators, and candidate quantum spin-liquids. Our work thus motivates the study and creation of interfaces between metals and Mott insulators, not only to provide routes to probe the spin correlations of the correlated subsystem, but also as a novel platform for stabilising and observing new physics.

**Materials and Methods**

Single-crystal samples of PdCrO$_2$ were grown by a NaCl-flux method in sealed quartz tubes as described in Ref. *(11)*. They were cleaved *in situ* at the measurement temperature of 6–13 K. High resolution ARPES measurements (Figs. 1C, 1D, 3A and Supplementary Fig. S1.) were performed at the I05 beamline of Diamond Light Source, UK, using a Scienta R4000 hemispherical electron analyser. The spectra shown in the main text were measured using p-polarised light, while fits to data taken using both s- and p-polarised light are included in Fig. 3A and Supplementary Fig. S5. The soft X-ray measurements (Figs. 2 and 4) were performed with p-polarised light at the I09 beamline of Diamond Light Source, UK; the ARPES measurements were performed using a Specs Phoibos 225 hemispherical electron analyser, while the X-ray absorption was recorded in the total electron yield mode, and is normalized by the photon flux. Further details of the theoretical methods are described in the Supplementary Text.

**Acknowledgments:**

We thank J. Schmalian, K. Kuroki, C. Hooley, A. Rost and B. Schmidt for useful discussions. We acknowledge support from the European Research Council (Grant Nos. ERC-714193-QUESTDO and ERC-319286-QMAC), the Royal Society, the Leverhulme Trust (Grant Nos. RL-2016-006 and PLP-2015-144R), the Max-Planck Society, The Simons Foundation, and the International Max-Planck Partnership for Measurement and Observation at the Quantum Limit. V.S. and O.J.C. acknowledge EPSRC for PhD studentship support through grant numbers EP/L015110/1 and EP/K503162/1. I.M. acknowledges PhD studentship support from the IMPRS for the Chemistry and Physics of Quantum Materials. We thank Diamond Light Source for access to beamlines I09 (Proposal No. SI19479) and I05 (Proposal No. SI17699), which contributed to the results presented here.

(a) **Author Contributions:** V.S., F.M., O.J.C., M.W., I.M., D.B. and P.D.C.K. measured the experimental data, and V.S. and F.M. analysed the data. S.Khim and P.K. grew and characterised the samples. V.S. performed initial tight binding modeling, S.Kitamura and T.O. the strong coupling theory calculations, H.R. the density functional theory calculations, and L.P. and A.G. the dynamical mean-field theory calculations. T.K.K., P.K.T., and T.-L.L. maintained the ARPES and soft x-ray ARPES end stations and provided experimental support. V.S., R.M., T.O., A.P.M. and P.D.C.K. wrote the manuscript with input and discussion from co-authors, and were responsible for overall project planning and direction.

(b) **Competing Interests:** The authors declare no competing interests.

**Supplementary Materials:**

Supplementary Text

Figures S1-S5

Tables S1-S3

# Supplementary Materials for

Probing spin correlations using angle resolved photoemission in a coupled metallic/Mott insulator system

V. Sunko[1,2]†, F. Mazzola[1]†, S. Kitamura[3]†, S. Khim[2], P. Kushwaha[2], O.J. Clark[1], M. Watson[1], I. Marković[1,2], D. Biswas[1], L. Pourovskii[4,5], T.K. Kim[6], T.-L. Lee[6], P.K. Thakur[6], H. Rosner[2], A. Georges[5], R. Moessner[3], T. Oka[3]*, A.P. Mackenzie[1,2]*, P.D.C. King[1]*

[1] SUPA, School of Physics and Astronomy, University of St. Andrews, St. Andrews KY16 9SS, United Kingdom.

[2] Max Planck Institute for Chemical Physics of Solids, Nöthnitzer Straße 40, 01187 Dresden, Germany.

[3] Max Planck Institute for the Physics of Complex Systems, Nöthnitzer Straße 38, 01187 Dresden, Germany.

[4] CPHT, Ecole Polytechnique, CNRS, Université Paris-Saclay, Route de Saclay, 91128 Palaiseau, France

[5] Institut de Physique, Collège de France, 75005 Paris, France.

[6] Diamond Light Source, Harwell Campus, Didcot, OX11 0DE, United Kingdom.

*Corresponding author. Email: oka@pks.mpg.de (T.O.), Andy.Mackenzie@cpfs.mpg.de (A.P.M.), philip.king@st-andrews.ac.uk (P.D.C.K.)

† These authors contributed equally to this work.

**This PDF file includes:**

    Supplementary Text
    Figures S1 to S5
    Tables S1 to S3

**Supplementary Text**

## 1) Density Functional Theory

Scalar-relativistic density functional theory (DFT) electronic structure calculations were performed using the full-potential FPLO code (27–29), version fplo18.00-52. For the exchange-correlation potential, within the local density approximation, the parameterization of Perdew-Wang (30) was chosen. To obtain precise band structure, the calculations were carried out on a well converged mesh of 8000 $k$-points (20x20x20 mesh, 781 points in the irreducible wedge of the Brillouin zone). For all calculations, the experimental crystal structure was used (13).

From the converged calculation, a 10-band tight-binding model (TBM) based on Wannier functions was constructed in an energy window between -3 eV and 2 eV. For the TBM, all Cr $3d$ and Pd $4d$ orbitals were taken into account. This results in very good agreement with the DFT bands in the relevant energy window near the Fermi level (see Supplementary Fig. S1). The extracted tight binding integrals and on-site energies (see Supplementary Table S1) were used as input parameters for the strong coupling theory (see Supplementary Text 3).

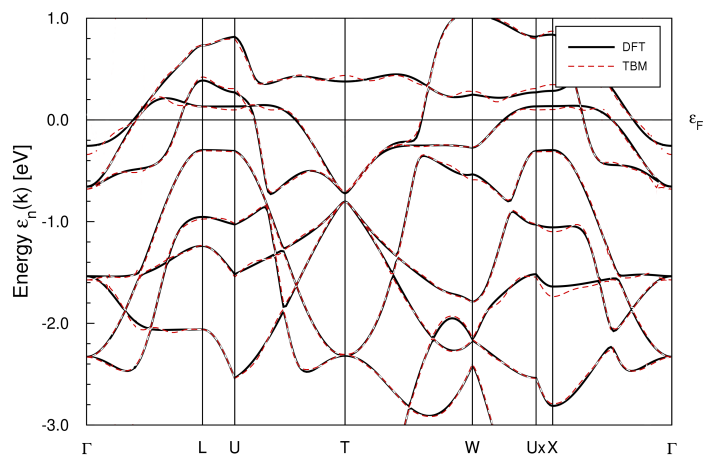

**Fig. S1.** Comparison of the DFT band structure (full black lines) and the Wannier function based 10-band tight-binding model (dashed red lines) near the Fermi level. Two ghost bands that have only weight for the unoccupied Cr $3d$ $e_g$ manifold have been removed for clarity.

## 2) DFT+DMFT calculations: spectral function

DFT + dynamical mean-field theory (DMFT) calculations for PdCrO$_2$ were carried for its experimental lattice structure (a=2.92 Å, c=18.09 Å) with a charge self-consistent framework (35, 36) combining the linearized augmented planewave band-structure code "Wien2k" (37) and the DMFT implementation provided by the library "TRIQS" (38, 39). Projective Wannier orbitals representing Cr $3d$ and Pd $4d$ states were constructed from Kohn-Sham bands in the energy range of [-5:11] eV. The rotationally-invariant on-site Coulomb repulsion vertex between Cr $3d$ orbitals was specified by the Slater parameter $F^0$ = 4.5 eV and Hund's rule coupling $J_H$ = 0.75 eV. We solved the DMFT quantum impurity problem employing the quasi-atomic Hubbard-I approximation (40). The double-counting correction was evaluated in the fully-localized limit using the atomic occupancy 3 of the Cr 3$d$ shell. Self-consistent DFT+DMFT calculations were converged to 0.05 mRy in the total energy.

The resulting total $k$-resolved DFT+DMFT spectral function $A(\mathbf{k}, \omega)$ is displayed in Supplementary Fig. S2. Within the Hubbard-I approximation the DMFT self-energy has no imaginary part, hence, there is also no lifetime broadening in our resulting $A(\mathbf{k}, \omega)$. Apart from this, however, our calculated spectral function is in excellent agreement with that obtained in Ref. *(20)* using a numerically exact quantum Monte Carlo method (one may notice a rather weak lifetime broadening for the occupied part of $A(\mathbf{k}, \omega)$ shown in Fig. 3 of Ref. *(20)*).

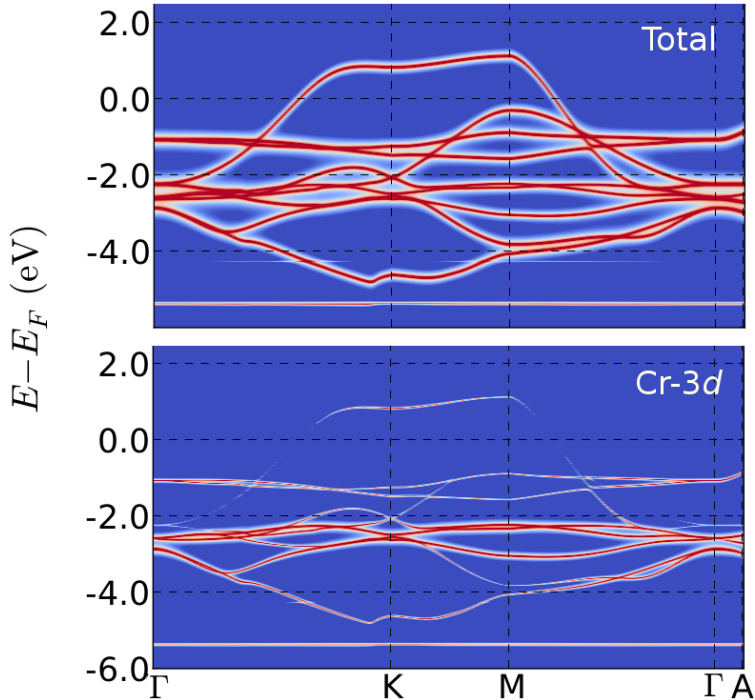

**Fig. S2.** Total (upper panel) and projected Cr 3$d$ (low panel) $k$-resolved DFT+DMFT spectral function $A(\mathbf{k}, \omega)$ of PdCrO$_2$.

The largest contribution of the Cr 3$d$ character to the DFT+DMFT $A(\mathbf{k}, \omega)$ is seen in the range from -2.5 to -2 eV, in good agreement with the location of diffuse spectral weight in the experimental on-resonance ARPES (Fig. 2B of the main text). There is almost no Cr contribution to the dispersive bands crossing $E_F$, confirming that they are of dominant Pd 4$d$ character. This finding is supported by our ARPES measurements: as well as a lack of spectral-weight enchancement of this band across the Cr L$_{2,3}$ – edge resonances ($I_{MB}$ in Fig. 2D of the main text), our high-resolution measurements indicate that its Fermi velocity is very similar in PdCrO$_2$ and the non-magnetic sister compound PdCoO$_2$, and very fast in both (Supplementary Fig. S3). This rules out any significant correlation-driven mass enhancement of these states.

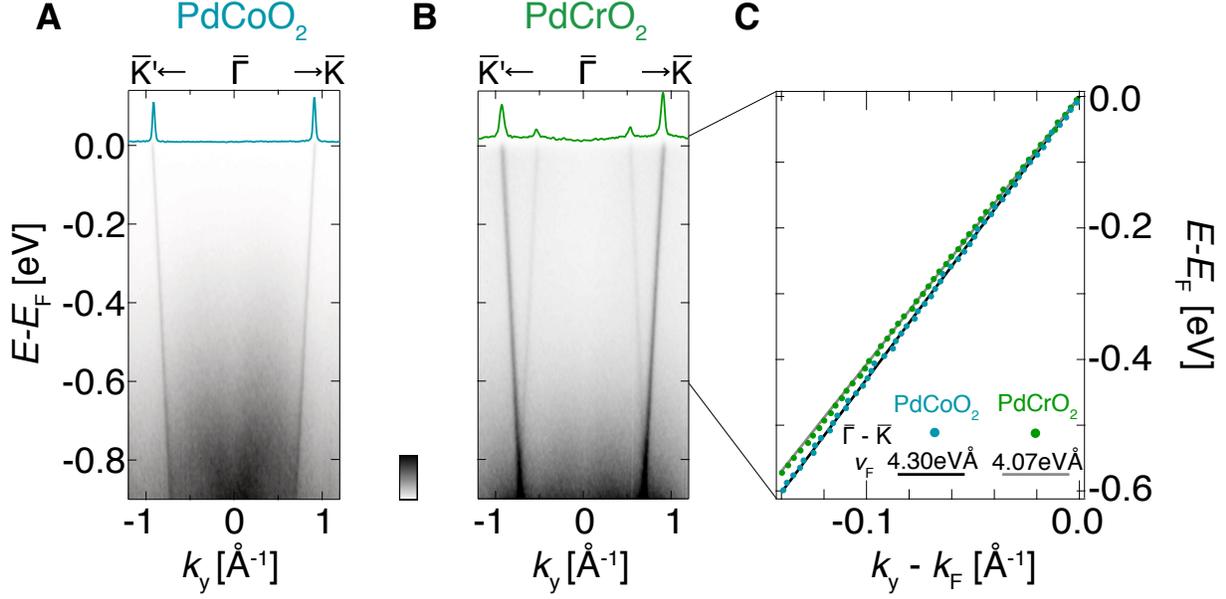

**Fig. S3.** Comparison of electronic structure in PdCoO$_2$ and PdCrO$_2$. Dispersion measured by ARPES ($hv = 110$ eV, $T = 6$ K) along the $\bar{\Gamma}\bar{K}$ direction of (**A**) PdCoO$_2$ and (**B**) PdCrO$_2$. (**C**) Fits to MDCs in the vicinity of the Fermi level for both compounds, showing fast Fermi velocities (linear fits). In PdCrO$_2$ the weak interplane coupling renormalises the Fermi velocity by at most 6% as compared to PdCoO$_2$, although a slight lattice expansion for PdCrO$_2$ vs. PdCoO$_2$ means that the actual degree of correlation-driven mass enhancement is likely even smaller than this value. This justifies the treatment of the Pd electrons as uncorrelated in the intertwined spin-charge model discussed in the main text.

### 3) Strong coupling theory

The demonstration of a Mott insulating state of the CrO$_2$ layers justifies the strong-coupling theory for the spectral function of the PdCrO$_2$ utilised here. While we show a simplified version without the orbital degree of freedom in the main text, here we employ a more realistic four-orbital model:

$$H = -\sum_{ij\sigma} t^p_{ij} p^\dagger_{i\sigma} p_{j\sigma} - \sum_{ijmm'\sigma} t^{mm'}_{ij} c^\dagger_{im\sigma} c_{jm'\sigma} + \sum_{ijm\sigma} \left( g^m_{ij} p^\dagger_{i\sigma} c_{jm\sigma} + g^{m*}_{ij} c^\dagger_{jm\sigma} p_{i\sigma} \right) + H_{\text{int}}, \quad (1)$$

where $p_{j\sigma}$ represents the $d_{3z^2-r^2}$ orbital of Pd electrons on the $j$th site, while $c_{jm\sigma}$ represents the $3d$ orbitals of Cr electrons. $m = 1,2,3$ labels the three low-energy states of the crystal field splitting,

$$|m=1\rangle = |d_{3z^2-r^2}\rangle, \qquad |m=2\rangle = \cos\phi\,|d_{x^2-y^2}\rangle + \sin\phi\,|d_{xz}\rangle,$$
$$|m=3\rangle = \sin\phi\,|d_{yz}\rangle - \cos\phi\,|d_{xy}\rangle \quad (2)$$

with $\tan\phi \sim 0.693$, which are schematically drawn in Supplementary Fig. S3. Pd and Cr respectively form triangular lattices and are stacked alternately. The hopping parameters are estimated from the 10-orbital (Pd $4d$ and Cr $3d$ electrons) model constructed from the first-principles calculations (see Supplementary Text 1), and are summarized in Supplementary Table S1. In reality, the metallic Pd band of the $d_{3z^2-r^2}$ orbital is substantially hybridized with $d_{xy}$ and $d_{x^2-y^2}$ orbitals. While the multi-orbital treatment is necessary for orbitally-resolved properties, it is not crucial for reproducing the parabolic energy spectrum and so we neglected the detailed orbital components of Pd electrons for simplicity. The derivation of the Kondo Hamiltonian and the associated spectral formula is streightforwardly applicable to the case where the Pd electrons have the orbital indices as well.

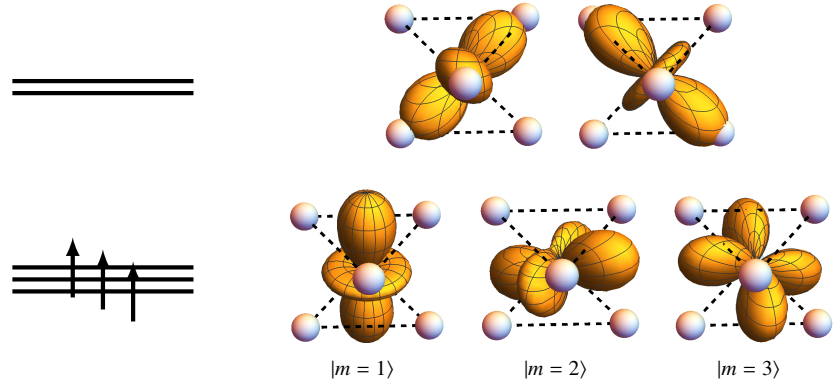

**Fig S4.** Schematic picture of the Cr $3d$ orbitals. The three lower energy states are labeled as $m = 1,2,3$.

We explicitly include the onsite repulsion $U$ and the Hund coupling $J_H$ for the Cr electrons as

$$H_{\text{int}} = \left(\frac{1}{2}U - \frac{5}{4}J_H\right)\sum_i \left[\sum_m (n_{im} - 1)\right]^2 + \frac{1}{2}J_H \sum_{im}(n_{im} - 1)^2 - J_H \sum_i \mathbf{S}_i^2 \quad (3)$$

to capture the correlated nature of the $3d$ electrons. Here $n_{im} = \sum_\sigma c^\dagger_{im\sigma} c_{im\sigma}$ and

$$\mathbf{S}_i = \frac{1}{2}\sum_{m\sigma\sigma'} c^\dagger_{im\sigma}\boldsymbol{\sigma}_{\sigma\sigma'} c_{im\sigma'} \quad (4)$$

is the localized spin of Cr.

| $\boldsymbol{R}_j - \boldsymbol{R}_i$ | $\pm\sqrt{3}(0,1,0)$ | $\pm\sqrt{3}(\frac{\sqrt{3}}{2},\frac{1}{2},0)$ | $\pm\sqrt{3}(\frac{\sqrt{3}}{2},-\frac{1}{2},0)$ | $\pm 3(1,0,0)$ | $\pm 3(\frac{1}{2},\frac{\sqrt{3}}{2},0)$ | $\pm 3(\frac{1}{2},-\frac{\sqrt{3}}{2},0)$ |
|---|---|---|---|---|---|---|
| $t_{ij}^p$ (meV) | 568 | | | $-108$ | | |

| $\boldsymbol{R}_j - \boldsymbol{R}_i$ | $\pm\sqrt{3}(0,1,0)$ | $\pm\sqrt{3}(\frac{\sqrt{3}}{2},\frac{1}{2},0)$ | $\pm\sqrt{3}(\frac{\sqrt{3}}{2},-\frac{1}{2},0)$ |
|---|---|---|---|
| $t_{ij}^{mm'}$ (meV) | $\begin{pmatrix} 141 & 0 & 152 \\ 0 & 17.6 & 0 \\ 152 & 0 & -44.9 \end{pmatrix}$ | $\begin{pmatrix} 48.4 & -53.4 & -76.1 \\ -53.4 & 110 & 132 \\ -76.1 & 132 & -44.9 \end{pmatrix}$ | $\begin{pmatrix} 48.4 & 53.4 & -76.1 \\ 53.4 & 110 & -132 \\ -76.1 & -132 & -44.9 \end{pmatrix}$ |

| $\boldsymbol{R}_j - \boldsymbol{R}_i$ | $\pm(-1,0,\frac{1}{2})$ | $\pm(\frac{1}{2},-\frac{\sqrt{3}}{2},\frac{1}{2})$ | $\pm(\frac{1}{2},\frac{\sqrt{3}}{2},\frac{1}{2})$ | $\pm(2,0,\frac{1}{2})$ | $\pm(-1,\sqrt{3},\frac{1}{2})$ | $\pm(-1,-\sqrt{3},\frac{1}{2})$ |
|---|---|---|---|---|---|---|
| $g_{ij}^m$ (meV) | $\begin{pmatrix} 110 \\ 0 \\ -52.6 \end{pmatrix}$ | $\begin{pmatrix} -55.3 \\ 95.7 \\ -52.6 \end{pmatrix}$ | $\begin{pmatrix} -55.3 \\ -95.7 \\ -52.6 \end{pmatrix}$ | $\begin{pmatrix} 122 \\ 0 \\ 122 \end{pmatrix}$ | $\begin{pmatrix} -60.8 \\ 105 \\ 122 \end{pmatrix}$ | $\begin{pmatrix} -60.8 \\ -105 \\ 122 \end{pmatrix}$ |

**Table S1.** Table of hopping parameters.

3.1 Derivation of the Kondo lattice Hamiltonian

Since the presence of the Mott transition of Cr electrons is demonstrated in the DFT+DMFT calculation (see Supplementary Text 2) we perform the strong coupling expansion to investigate the low-energy structure of the Mott insulating phase. We derive an effective Hamiltonian by applying the Schrieffer-Wolff transformation,

$$H_{\text{eff}} = e^{iS}He^{-iS} = H + [iS,H] + \frac{1}{2}[iS,[iS,H]] + \cdots, \quad (5)$$

where $S$ is determined such that terms changing the number of doublon, holon, or $\boldsymbol{S}_i^2 = S_i(S_i + 1)$ do not appear in the effective Hamiltonian. The Schrieffer-Wolff transformation thus 'eliminates' the high energy charge (doublon-holon) excitation of the Cr atoms. The upper and lower Hubbard bands, the latter of which is visible in experiment (Fig2(B,C) of the main text) is thus missing in the Kondo-Heisenberg model. However, all excitations involving the spin degree of freedom are kept in the model after the transformation. We consider a perturbative expansion from the atomic limit and construct $S$ order by order: We expand

$$S = S^{(1)} + S^{(2)} + \cdots$$

and compare each order, where hopping terms are allocated to the first order and interacting terms the zeroth order. From this condition, the first-order term $S^{(1)}$ must satisfy

$$iS^{(1)}|GS\rangle = \frac{1}{U_{\text{eff}}} \sum_{ijm\sigma} \left[ 2\left(g_{ij}^m p_{i\sigma}^\dagger c_{jm\sigma} + g_{ij}^{m*} c_{jm\sigma}^\dagger p_{i\sigma}\right) - \sum_{m'} t_{ij}^{mm'} c_{im\sigma}^\dagger c_{jm'\sigma} \right] |GS\rangle, \quad (6)$$

where $|GS\rangle$ is the (macroscopically-degenerate) ground state (no doublons/holons, $S = 3/2$ on each site) of $H_{\text{int}}$ and $U_{\text{eff}} = U + 2J_H$. We obtain the second-order effective Hamiltonian (up to constants) from this relation as

$$H_{\text{eff}} = -\sum_{ij\sigma} t_{ij}^p p_{i\sigma}^\dagger p_{j\sigma} + \frac{1}{2}\sum_{ij} J_{ij}\boldsymbol{S}_i \cdot \boldsymbol{S}_j + \sum_{ijk\sigma\sigma'} K_{ijk} p_{i\sigma}^\dagger (\boldsymbol{S}_j \cdot \boldsymbol{\sigma}_{\sigma\sigma'}) p_{k\sigma'} \quad (7)$$

with

$$J_{ij} = \sum_{mm'} \frac{4|t_{ij}^{mm'}|^2}{9U_{\text{eff}}}, \qquad K_{ijk} = \sum_{m} \frac{4g_{ij}^{m} g_{kj}^{m*}}{3U_{\text{eff}}}. \qquad (8)$$

The obtained form is identical to that of the simplified version (Equation 2 of the main text) but with $S = 3/2$ rather than $S = 1/2$. The last term can be thought as the Kondo coupling between the local Cr spin $S_j$ and neighbouring Pd electrons. Representative values are determined from the DFT calculations, and are listed in Table S2.

| $R_j - R_i$ | $(\frac{3}{2}, \frac{\sqrt{3}}{2}, 0)$ | $(-\frac{1}{2}, -\frac{\sqrt{3}}{2}, 1)$ | $(-2, 0, 1)$ |
|---|---|---|---|
| $J_{ij}$ (meV) | 5.17 | 0.166 | 0.106 |

| $R_j - R_i$ | $(-1, 0, \frac{1}{2})$ | $(2, 0, \frac{1}{2})$ | $(2, \sqrt{3}, \frac{1}{2})$ |
|---|---|---|---|
| $K_{iji}$ (meV) | 3.39 | 6.71 | 1.19 |

**Table S2**. Table of spin coupling constants derived from the strong coupling expansion.

These values of superexchange interactions $J_{ij}$ between Cr $S = 3/2$ spins are in good agreement with those evaluated from the converged DFT+DMFT results with the linear-response theory of Ref. (*41*) for several first correlation shells (Supplementary Fig. S5). The obtained values are listed in Supplementary Table S3.

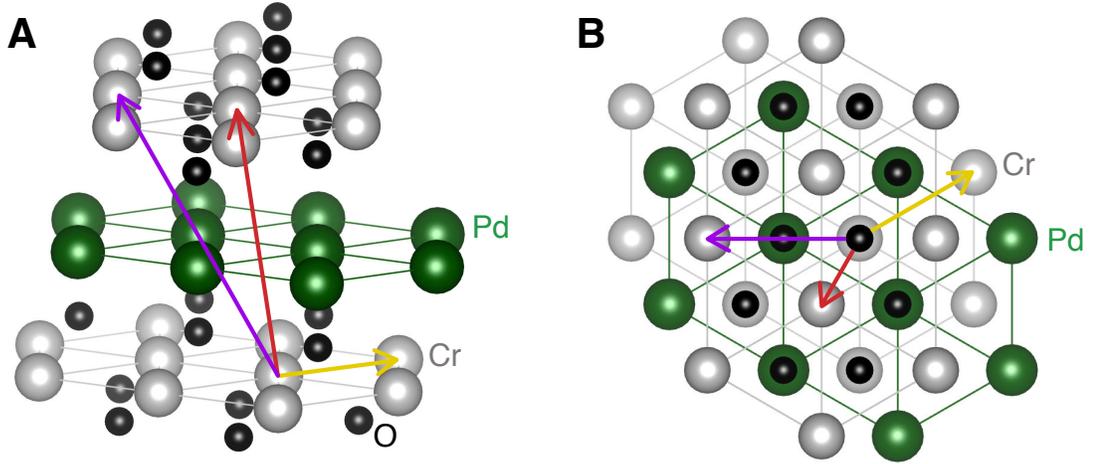

**Fig. S5.** Exchange pathways in PdCrO$_2$. A side view (**A**) and a view from above (**B**) of the crystal structure of PdCrO$_2$, with superexchange interactions marked (see Tables S2 and S3). Yellow, red and purple arrows indicate the lattice vectors $R_i - R_j = \left(\frac{3}{2}, \frac{\sqrt{3}}{2}, 0\right), \left(-\frac{1}{2}, -\frac{\sqrt{3}}{2}, 1\right)$ and $(-2, 0, 1)$ respectively.

| $R_j - R_i$ | $(\frac{3}{2}, \frac{\sqrt{3}}{2}, 0)$ | $(-\frac{1}{2}, -\frac{\sqrt{3}}{2}, 1)$ | $(-2, 0, 1)$ |
|---|---|---|---|
| $J_{ij}$ (meV) | 6.24 | −0.06 | 0.12 |

**Table S3.** Cr-Cr superexchange interactions $J_{ij}$ (in meV) calculated by the DFT+DMFT method for the lattice vectors indicated in Fig. S4.

### 3.2 Spectral function

The spectral function of Cr electrons is represented as

$$A_{\text{Cr}}(\mathbf{k}, \omega) = \sum_{m\sigma\alpha\alpha'} \left| \sum_i \langle \alpha_0^{(N)} | c_{im\sigma} | \alpha_0'^{(N+1)} \rangle e^{-i\mathbf{k}\mathbf{r}_i} \right|^2 \delta\left(\omega + E_\alpha^{(N)} - E_{\alpha'}^{(N+1)}\right) \frac{e^{-\beta E_\alpha^{(N)}}}{Z} (1 + e^{-\beta\omega}), \quad (9)$$

where $\left|\alpha_0^{(N)}\right\rangle$ is the eigenstate of the original Hamiltonian with $N$ electrons and eigenenergy $E_\alpha^{(N)}$. We rewrite this expression in terms of the eigenstate of the effective Hamiltonian which is given as $\left|\alpha^{(N)}\right\rangle = e^{iS} \left|\alpha_0^{(N)}\right\rangle$.

We can neglect terms with $E_\alpha^{(N)} = \mathcal{O}(U_{\text{eff}})$ since $e^{-\beta E_\alpha^{(N)}} \ll 1$ holds for the temperature range of interest. Terms with $E_{\alpha'}^{(N+1)} = \mathcal{O}(U_{\text{eff}})$ can also be neglected for the low-energy regime $|\omega| = \left|E_\alpha^{(N)} - E_{\alpha'}^{(N+1)}\right| \ll U_{\text{eff}}$. Then the leading order expression in $1/U_{\text{eff}}$ is obtained as

$$A_{\text{Cr}}(\mathbf{k}, \omega) \sim \sum_{m\sigma\alpha\alpha'} \left| \sum_i \langle \alpha^{(N)} | [iS^{(1)}, c_{im\sigma}] | \alpha'^{(N+1)} \rangle e^{-i\mathbf{k}\mathbf{r}_i} \right|^2 \delta\left(\omega + E_\alpha^{(N)} - E_{\alpha'}^{(N+1)}\right) \frac{e^{-\beta E_\alpha^{(N)}}}{Z} (1 + e^{-\beta\omega}) \quad (10)$$

$$\sim \sum_{m\sigma\alpha\alpha'} \left| \int \frac{d^3 q}{(2\pi)^3} \sum_{\sigma'} \frac{4 g_{k+q}^{m*}}{3U_{\text{eff}}} \langle \alpha^{(N)} | \mathbf{S}_{-\mathbf{q}} \cdot \boldsymbol{\sigma}_{\sigma\sigma'} p_{\mathbf{k}+\mathbf{q}\sigma'} | \alpha'^{(N+1)} \rangle \right|^2 \delta\left(\omega + E_\alpha^{(N)} - E_{\alpha'}^{(N+1)}\right) \frac{e^{-\beta E_\alpha^{(N)}}}{Z} (1 + e^{-\beta\omega}), \quad (11)$$

where $\mathbf{S}_\mathbf{q} = \sum_i \mathbf{S}_i e^{-i\mathbf{q}\cdot\mathbf{r}_i}$ and $g_\mathbf{k}^m = \sum_j g_{ij}^m e^{-i\mathbf{k}\cdot(\mathbf{r}_i - \mathbf{r}_j)}$. This expression can be equivalently obtained from the correlation function

$$G_{\text{Cr}}^{\text{eff}}(\mathbf{k}, t) = -i \sum_{m\sigma\sigma'\sigma''} \int \frac{d^3 q}{(2\pi)^3} \int \frac{d^3 q'}{(2\pi)^3} \frac{16 g_{\mathbf{k}+\mathbf{q}}^{m*} g_{\mathbf{k}+\mathbf{q}'}^m}{9 U_{\text{eff}}^2} \langle \{\mathbf{S}_{-\mathbf{q}}(t) \cdot \boldsymbol{\sigma}_{\sigma\sigma'} p_{\mathbf{k}+\mathbf{q}\sigma'}(t), \mathbf{S}_{\mathbf{q}'} \cdot \boldsymbol{\sigma}_{\sigma''\sigma} p_{\mathbf{k}+\mathbf{q}'\sigma''}^\dagger \} \rangle \quad (12)$$

with $O(t)$ being the Heisenberg representation (for the effective Hamiltonian) of an operator $O$. Namely, the low-energy excitation of the Cr electrons is described by a simultaneous disturbance on the metallic layer and the localized spin.

Since the coupling between the metallic and insulating layer is small ($\sim |g|^2/U_{\text{eff}}$), we can decouple the expectation value $\langle SSpp^\dagger\rangle \sim \langle SS\rangle \langle pp^\dagger\rangle$ in the leading-order evaluation. This treatment is justified when the vertex correction (in terms of the diagrammatic expansion about the inter-layer coupling) can be neglected. The correlation function can then be approximated as

$$G_{\text{Cr}}^{\text{eff}}(\boldsymbol{k}, t) \sim -i \sum_{m\sigma} \int \frac{d^3\boldsymbol{q}}{(2\pi)^3} \frac{16|g_{\boldsymbol{k}+\boldsymbol{q}}^m|^2}{9 U_{\text{eff}}^2} [\langle \boldsymbol{S}_{-\boldsymbol{q}}(t) \cdot \boldsymbol{S}_{\boldsymbol{q}}\rangle \langle p_{\boldsymbol{k}+\boldsymbol{q}\sigma}(t) p_{\boldsymbol{k}+\boldsymbol{q}\sigma}^\dagger\rangle$$
$$+ \langle \boldsymbol{S}_{\boldsymbol{q}} \cdot \boldsymbol{S}_{-\boldsymbol{q}}(t)\rangle \langle p_{\boldsymbol{k}+\boldsymbol{q}\sigma}^\dagger p_{\boldsymbol{k}+\boldsymbol{q}\sigma}(t)\rangle] \quad (13)$$

where we neglect the off-diagonal element $\langle p_{\boldsymbol{k}+\boldsymbol{q},\sigma}(t) p_{\boldsymbol{k}+\boldsymbol{q}',\sigma'}^\dagger\rangle$ which can exist in the ordered phase. Equation (3) in the main text is obtained by considering the zero temperature limit of the Fourier-transform of the expression for $G_{\text{Cr}}^{\text{eff}}(\boldsymbol{k}, t)$.

When the spins acquire a long-range order, the spin correlation function $\langle \boldsymbol{S}_{\boldsymbol{q}} \cdot \boldsymbol{S}_{-\boldsymbol{q}}(\omega)\rangle = \int dt \langle \boldsymbol{S}_{\boldsymbol{q}} \cdot \boldsymbol{S}_{-\boldsymbol{q}}(t)\rangle e^{-i\omega t}$ has a divergent peak at $\omega = 0$ and characteristic values of $\boldsymbol{q}$. For the present case, the experimentally determined magnetic order *(13, 14)* has a 120-degree structure, i.e.,

$$\langle \boldsymbol{S}_i\rangle = S \begin{pmatrix} \sin(\boldsymbol{Q}+\boldsymbol{Z})\cdot \boldsymbol{r}_i \\ \cos \boldsymbol{Q}\cdot \boldsymbol{r}_i \\ 0 \end{pmatrix} \quad (14)$$

with $\boldsymbol{Q} = (2\pi/3, 2\pi/3, 0)$ and $\boldsymbol{Z} = (0, 0, \pi)$. With this magnetic order, we obtain the back-folded steep bands as shown in Fig. 3E of the main text.

The spectral intensity of the back-folded band with the momentum shift of $\pm \boldsymbol{Q}$ can be approximated as

$$I_{\text{Cr}}^{\pm \boldsymbol{Q}}(\boldsymbol{k}) = \sum_m \frac{8S^2 |g_{\boldsymbol{k}\pm\boldsymbol{Q}}^m|^2}{9 U_{\text{eff}}^2}, \quad (15)$$

(fits to the calculated Cr spectral function are plotted as a function of energy in Fig. 3A of the main text (solid line)). This form of the intensity is substantially different from that of the back-folded band of Pd electrons due to the modified potential.

If we consider a bilayer case for simplicity, the 'band folding' model Pd Hamiltonian, with the modified potential due to the 120-degree structure, can be written in a $2 \times 2$ form:

$$H = \sum_{\boldsymbol{k}} \begin{pmatrix} p_{\boldsymbol{k}\uparrow} \\ p_{\boldsymbol{k}+\boldsymbol{Q}\downarrow} \end{pmatrix}^\dagger \begin{pmatrix} \epsilon_{\boldsymbol{k}} & \Delta \\ \Delta & \epsilon_{\boldsymbol{k}+\boldsymbol{Q}} \end{pmatrix} \begin{pmatrix} p_{\boldsymbol{k}\uparrow} \\ p_{\boldsymbol{k}+\boldsymbol{Q}\downarrow} \end{pmatrix}. \quad (16)$$

The spectral intensity of the back-folded band with the momentum shift of $\pm \boldsymbol{Q}$ is obtained as

$$I_{\text{Pd}}(\boldsymbol{k}) = 1 - \frac{|\epsilon_{\boldsymbol{k}} - \epsilon_{\boldsymbol{k}+\boldsymbol{Q}}|}{\sqrt{(\epsilon_{\boldsymbol{k}} - \epsilon_{\boldsymbol{k}+\boldsymbol{Q}})^2 + 4\Delta^2}}, \quad (17)$$

which rapidly decays as $I_{\text{Pd}}(\boldsymbol{k}) \sim 2|\Delta/(\epsilon_{\boldsymbol{k}} - \epsilon_{\boldsymbol{k}+\boldsymbol{Q}})|^2$ as $|\epsilon_{\boldsymbol{k}} - \epsilon_{\boldsymbol{k}+\boldsymbol{Q}}|$ becomes large. We plot this form of intensity in Fig. 3A of the main text (dashed line), with $\Delta = 20$ meV, consistent with the experimental gap $2\Delta \sim 40$ meV estimated from the breakdown field reported in Ref. 16 as $\hbar \omega_c \sim \Delta^2 / E_F$. We stress that, for such a "conventional" back-folding, the rapid suppression of spectral weight away from the magnetic Brillouin zone boundary is therefore intrinsic. In contrast,

in the intertwined spin-charge model, the reconstructed weight in the Cr spectral function is approximately constant, varying only due to momentum-dependent variations of the inter-layer coupling term $g_{k+Q}$. Our calculations (Supplementary Fig. S6) show how this is sensitive to details of the calculation. Including only nearest-neighbour coupling, the spectral weight slightly decreases towards the Fermi level. Including next nearest-neighbour coupling, the spectral weight increases towards the Fermi level. In all cases, the variation in spectral weight is less than a factor of 2 over an energy range of more than 700 meV below the Fermi level, entirely consistent with our experimental measurements shown in Fig. 3A of the main text and Supplementary Fig. S6. Throughout, we have shown the relative intensity variations between different models and the experimental data by normalising to the intensity at -0.7 eV binding energy; equivalent conclusions are drawn if normalising directly by the main band intensity, as shown in Supplemental Fig. S7.

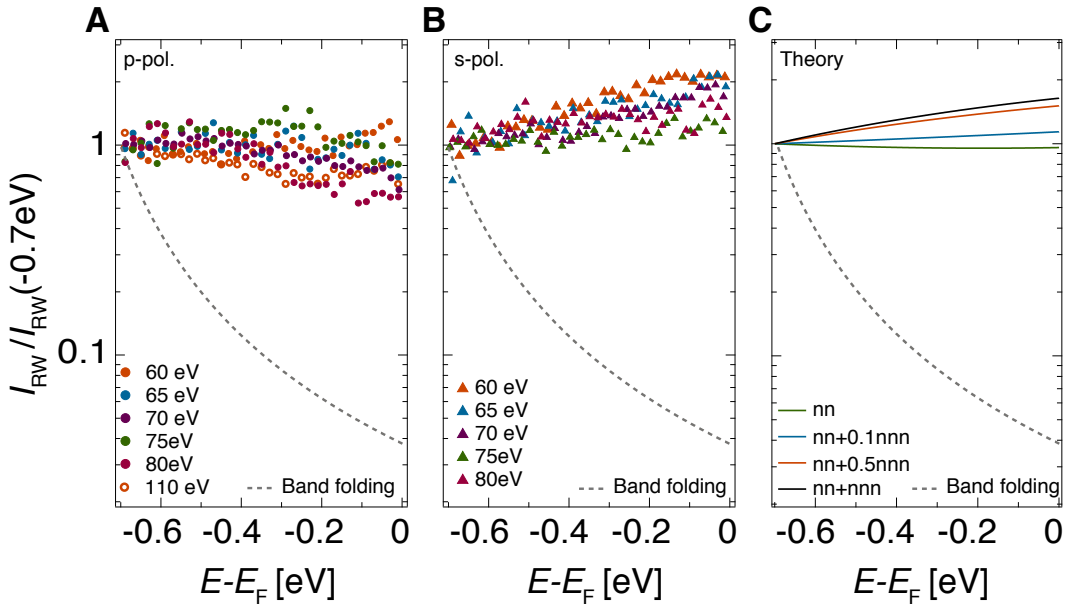

**Fig. S6.** Binding-energy dependent spectral weight. Reconstructed weight ($I_{RW}$) as obtained from fits to dispersions measured using various photon energies and (**A**) p-polarised and (**B**) s-polarised light. Although varying ARPES matrix elements cause some changes in the binding energy dependence of the reconstructed weight intensity for the different measurement conditions, the measured spectral weight never varies by more than a factor of two over the 700meV energy range. This is in sharp contrast to the prediction of the simple 'band folding' model. While binding energy independent 'shadow features' have previously been observed in materials with superperiodic structures or structural distortions *(38)*, such underlying origins would be inconsistent with the Cr character of the reconstructed weight that we observe. (**C**) Cr $I_{RW}$ predicted by the intertwined spin-charge model for various combinations of nearest neighbour (nn) and next-nearest neighbour (nnn) hopping parameters. Small parameter-dependent quantitative variations are observed, although the overall binding energy dependence of the reconstructed weight intensity remains weak for all parameters, again in clear contrast to the 'band folding' model.

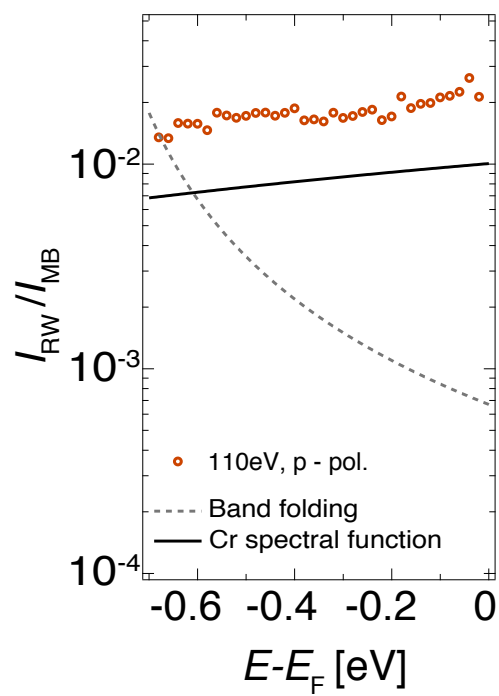

**Fig. S7.** The ratio of the reconstructed weight and the main band weight found experimentally (symbols), for the 'band folding' model (dashed line) and the Cr spectral function predicted by our theory (full line). Experimental data are divided by the calculated ratio of photoemission cross-sections of Cr 3d and Pd 4d orbitals, which is 11 at 110eV.